\documentclass{pasj00}
\draft

\begin{document}
\SetRunningHead{Kaneda et al.}{AKARI telescope}
\Received{//}%{yyyy/mm/dd}
\Accepted{//}%{yyyy/mm/dd}

\title{Far-infrared distributions in nearby spiral galaxies NGC2841 and NGC2976 observed with AKARI/FIS}

%%% begin:list of authors
\author{%
   Hidehiro \textsc{Kaneda},\altaffilmark{1}
   Toyoaki \textsc{Suzuki},\altaffilmark{1}
   Takashi \textsc{Onaka},\altaffilmark{2}
   Yasuo \textsc{Doi},\altaffilmark{3}\\
   Mitsunobu \textsc{Kawada},\altaffilmark{4}
   Bon-Chul \textsc{Koo},\altaffilmark{5}
   Sin'itirou \textsc{Makiuti},\altaffilmark{1}
   Takao \textsc{Nakagawa},\altaffilmark{1}\\
   Yoko \textsc{Okada},\altaffilmark{1}
   Stephen \textsc{Serjeant},\altaffilmark{6}
   Hiroshi \textsc{Shibai},\altaffilmark{4}
   and
   Mai \textsc{Shirahata},\altaffilmark{1}}

 \altaffiltext{1}{Institute of Space and Astronautical Science, \\
Japan Aerospace Exploration Agency, Sagamihara, Kanagawa 229-8510}
 \email{kaneda@ir.isas.jaxa.jp}
 \altaffiltext{2}{Department of Astronomy, Graduate School of Science, University of Tokyo, \\
Bunkyo-ku, Tokyo 113-0003}
 \altaffiltext{3}{Department of Earth Science and Astronomy, University of Tokyo, Meguro-ku, Tokyo 153-8902}
 \altaffiltext{4}{Graduate School of Science, Nagoya University, Chikusa-ku, Nagoya 464-8602}
 \altaffiltext{5}{Department of Physics and Astronomy, Seoul National University, Seoul 151-742, KOREA}
 \altaffiltext{6}{Department of Physics, The Open University, Milton Keynes, MK7 6AA, ENGLAND}

%%% end:list of authors

%% `\KeyWords{}' always has to be placed before `\maketitle'.
\KeyWords{dust, extinction --- infrared: galaxies --- galaxies: individual (NGC~2841, NGC~2976) --- galaxies: spirals --- galaxies: ISM} %Do NOT move this preamble from here!

\maketitle

\begin{abstract}
Far-infrared (IR) images of the nearby Sb galaxy NGC~2841 and the Sc galaxy NGC~2976 at wavelengths of 65, 90, 140, and 160 $\mu$m have been obtained with the Far-Infrared Surveyor (FIS) onboard AKARI. Both galaxies reveal similar morphologies of dust rings. They are, however, significantly different in the dust temperature: a cold ($\sim$ 21 K) ring for NGC~2841 and a warm ($\sim$ 30 K) ring for NGC~2976, which presumably reflects the difference in the origin of the ring structure for the different Hubble type of the galaxy. In addition to the dust ring structure, a warm dust component is detected from the central region of NGC~2841, which may be attributed to the heating by its Low-Ionization Nuclear Emission-line Region nucleus. As for NGC~2976, an extended dust component is observed along the minor axis, which shows a distribution somewhat asymmetrical to the galactic disk; this might be associated with the HI bridge in the M81/M82 group that NGC~2976 belongs to. By taking advantage of a wealth of the far-IR bands of the FIS, it is demonstrated that the spectral energy distribution of NGC~2841 is spatially more variable than that of NGC~2976. 
 
\end{abstract}

\section{Introduction}
Nearby regular Sb and Sc galaxies provide us with valuable laboratories for understanding the large-scale structures of normal spiral galaxies like our own. Among them, the origin of the unusual gas distribution in our Galaxy such as a circular cloud ring at $R=4-8$ kpc (e.g. Scoville and Solomon 1975) is one of the issues yet to be completely understood. The ring-shaped concentration of gas at intermediate radii, or equivalently the lack of gas toward the center of a galaxy, could be linked to the presence of significant nuclear bulges; the gas originally at small radii may have been consumed up to form stars in the central bulge (Young and Scoville 1982b). Another explanation is that the ring cloud could arise from dynamical action of internal resonances, which can cause three types of rings: nuclear, inner, and outer ones (Young and Scoville 1982b; Buta 1986). The ring could also be caused by the high probability of the growth of two fluid instabilities in this region (Jog and Solomon 1984), or galaxy interactions like those seen in the Cartwheel galaxy (Hernquist and Weil 1993). Such gaseous rings are not universally present in spiral galaxies; indeed, a significant fraction of late-type galaxies exhibit CO distributions unlike that in our Galaxy (Young and Scoville 1982a, and references therein). Among tracers of the large-scale distribution of the interstellar medium (ISM), optical emission is overwhelmed by a bright stellar bulge, while HI and CO, which would be easily seen through the bulge, are largely absent in ionized gas regions like those near the center of M31 (Devereux et al. 1994). The far-infrared (IR) emission can penetrate the stellar bulge and trace all phases of the ISM, as recently demonstrated in M31 with Spitzer/MIPS (Gordon et al. 2006), complementing fundamental information about the distribution of mass in galaxies provided by rotation curves (e.g. Sofue and Rubin 2001). A dust ring, if present, can be used for studying the relation between dust emission and star formation activity.

Here we present new far-IR images of the nearby Sb galaxy NGC~2841 and Sc galaxy NGC~2976 taken with the Far-Infrared Surveyor (FIS; Kawada et al. 2007) onboard AKARI (Murakami et al. 2007). The properties of the observed galaxies are summarized in table 1; the distances to NGC~2841 and NGC~2976 are assumed to be 7.6 Mpc (Osman et al. 1982) and 3.6 Mpc (Karachentsev et al. 2002), respectively.  It is confirmed from table 1 that both spiral galaxies are similar in apparent optical sizes, inclination, and position angles, and thus an ideal combination for comparative research between the different Hubble types. NGC~2841 is classified as an isolated regular spiral without global spiral patterns (Kormendy and Norman 1979), which is known to have a weak Low-Ionization Nuclear Emission-line Region (LINER) nucleus (Hummel and Bosma 1982; Ho et al. 1995). 
The HI rotation curves show evidence for the presence of a massive nuclear bulge (Bosma 1978). The galaxy has a prominent molecular cloud ring; the CO map of Young and Scoville (1982b) shows a ring structure at radii of 2-6 kpc. The molecular ring is found to have bluer B-J and B-K colors than the central regions and the outer disk (Prieto et al. 1985), indicating more recent star formation in a younger population as would be expected in a region of higher gas density. Furthermore, NGC~2841 is reported to have a chemically and dynamically decoupled nucleus as well as an unusual counterrotating component of gas and stars in the outskirts of the nucleus, which gives rise to an idea that the galaxy experienced an encounter with a smaller gas-rich galaxy in the long past (Sil'chenko et al. 1997).  

NGC~2976 is a bulgeless, unbarred pure disk system, which is known to have a dark matter halo with a shallow density profile (Simon et al. 2003). This Sc galaxy is a member of the M81/M82 group of galaxies. The comparatively small mass and luminosity of NGC~2976 (see table 1) place it generically nearer to the family of dwarf galaxies rather than to normal galaxies (Bronkalla et al. 1992). The galaxy also has a possible molecular ring structure (Brouillet et al. 1992), which does not however present such a strong contrast against the rest of the galaxy as that in NGC~2841. The inner part of NGC~2976 contains a large number of young concentrated star forming regions; most of them are conspicuous H$\alpha$ regions (Hodge and Kennicutt 1983). The most active star formation is loosely concentrated at about 1.2 kpc distance from the center, possibly part of the ring. It is claimed that the star formation activity is triggered by infall of intergalactic matter (Bronkalla et al. 1992), which is supported by the existence of an HI-bridge that connects NGC~2976 to the huge HI cloud in the M81/82 group (Appleton et al. 1981)

NGC~2841 and NGC~2976 are among the targets of the Spitzer Infrared Nearby Galaxies Survey (SINGS) legacy program (Kennicutt et al. 2003; Dale et al. 2005); far-IR distributions of these galaxies are not yet reported in any paper. Due to the difference in the aperture size of the telescope, which is 85 cm for Spitzer and 68.5 cm for AKARI, the spatial resolution of the FIS is a little poorer than that of the MIPS; the full widths at the half maxima of the point spread functions (PSFs) measured in the orbit are $37''-61''$ at the 4 bands (Kawada et al. 2007), which correspond to linear scales of $1.3-2.2$ kpc on NGC~2841 and $0.6-1.1$ kpc on NGC~2976. Nevertheless, the 4 photometric far-IR bands of the FIS have a great advantage over the MIPS 2 far-IR bands to unambiguously determine the temperature and the mass of dust grains. We below report the nature of the far-IR distributions obtained from these nearby bright galaxies. Although the rather high inclination of these galaxies makes it difficult to determine their detailed ISM distributions, large-scale structures such as a dust ring and a dust flow from/into a disk are well observable in far-IR, with even higher singal-to-noise ratios owing to larger optical depths than the face-on configuration.

\section{Observations and Data Reduction}
The FIS observations of the galaxies NGC~2841 and NGC~2974 were carried out on April 26 and April 20 in 2006, respectively, as pointed observations during the AKARI performance verification phase. A summary of the observation log is listed in table 2. For each galaxy, we obtained N60 (centered at a wavelength of 65 $\mu$m with a width of 22 $\mu$m; Kawada et al. 2007), WIDE-S (90 $\mu$m with 38 $\mu$m), WIDE-L (140 $\mu$m with 52 $\mu$m), and N160 (160 $\mu$m with 34 $\mu$m) band images using one of the FIS observation modes, FIS01, a 2-round-trip slow scan with a scan speed of $8''$ sec$^{-1}$. A region approximately $10'\times 10'$ around each galaxy was covered with a scan map. The signal integration time is set to be as short as 0.5 sec for both observations to safely avoid any saturation effects. The original purpose of the observations was to evaluate the imaging performance of the FIS for extended bright sources in a slow scan observation mode with the short integration time. 

The far-IR images were processed from the FIS Time Series Data (TSD) by using the AKARI official pipeline as of December in 2006 being developed by the AKARI data reduction team. The images were further cleaned by removing the after-effects of cosmic-ray hitting and the latency of the illumination of the internal calibration lamps with the algorithm developed in Suzuki (2007), where the subtraction of the dark level and the correction for the variations of detector responsivity were also performed; the dark level was estimated from the data taken in the beginning and at the end of the observation when the internal shutter of the FIS was closed. The variation of the detector responsivity was obtained by low-order polynomial fitting to the output data for the calibration lamp switched on every minute during the observation. The background levels were estimated from nearby blank skies and subtracted from the images. The PSFs are reported to be a little blurred along a scan direction at signal levels lower than $5 \sim 20$ \% of the peaks, depending on the band (Kawada et al. 2007); we do not discuss about such low signal levels below.
\section{Results}
\subsection{Far-infrared images}
The far-IR 4-band images obtained for NGC~2841 and NGC~2976 are shown in figures 1 and 2, respectively, where the bin size for the N60 and WIDE-S maps is set to be 15 arcsec and that for the WIDE-L and N160 maps is 25 arcsec; both are about half the beam sizes. For every map, the image size is approximately $6'\times 6'$, and the contours are drawn from 25 \% to 95 \% of the background-subtracted peak surface brightness with the linear spacing of 10 \%. A rather conservative value of 25 \% for the lowest contour level is adopted by taking into account that the contour shapes at lower levels are more sensitive to uncertainties existing at the current stage of FIS calibration. As seen in the figure, the images show elongation in the far-IR emission aligned to the position angle of each galaxy, which is similar to each other (table 1). In all the bands except the N60 band, both galaxies show similar morphologies consistent with ring structures: the peak of emission is not located in the optical center but near the edges of the disk. In the N160 and WIDE-L bands, the peak surface brightness show similar values between the galaxies, whereas, in the WIDE-S band, NGC~2841 exhibits significantly lower peak surface brightness than NGC~2976. The N60 band images reveal a striking difference in the structure between the galaxies; NGC~2841 does not exhibit the ring structure any more, but shows a concentration of emission in the central region. The far-IR emission in the nuclear region has a component with warmer dust than the ring structure of NGC~2841, as the flux from the ring fades toward shorter wavelengths. This is consistent with the gas in the nuclear region being heated by the LINER nucleus, as well as emission from asymptotic giant branch stars in the bulge. In addition, there is a hint of the far-IR emission protruding from the disk toward the northeast in the maps of NGC~2976 (figure 2). This is not likely due to the artifact of the far-IR detectors, since there is no such possible emission seen in the maps of NGC~2841 when the maps of both galaxies are created from slow scan observations of similar scan directions between the north-northeast and the south-southwest. This might be associated with the HI bridge that runs almost in parallel to the minor axis and connects NGC~2976 to the huge intergalactic HI cloud in the M81/82 group (Appleton et al. 1981). 
\subsection{Spectral energy distributions}
We have derived the flux densities of NGC~2841 and NGC~2976 in the N60, WIDE-S, WIDE-L, and N160 bands of the FIS (table 3). These values are obtained by integrating the surface brightness within a radius of $3'$ from the center of each galaxy. The aperture sizes are shown by the large dashed circles in figures 1 and 2 for NGC~2841 and NGC~2976, respectively. The aperture is so large that the loss of fluxes outside the aperture can be neglected. Color corrections are performed for the 4 bands by using correction tables provided by the FIS team. The absolute fluxes are calibrated by using far-IR-bright asteroids (Kawada et al. 2007). The flux uncertainties including both systematic effects associated with the far-IR detectors and absolute uncertainties are estimated to be no more than 20 \% for N60 and WIDE-S, 30 \% for WIDE-L, and 40 \% for N160. The flux densities thus obtained from the FIS show excellent agreement with previous far-IR measurements (table 3; Dale et al. 2005; Rice et al. 1988; Sanders et al. 2003). The far-IR spectral energy distributions (SEDs) constructed from the flux densities in table 3 are presented in figure 3. It is obvious from these SEDs that NGC~2841 has significantly colder dust emission than NGC~2976. 
\section{Discussion}
\subsection{Dust temperature and mass}
We have obtained dust temperatures from the SEDs of NGC~2841 and NGC~2976 (figure 3) under the assumption that the far-IR emission of the galaxies originates from dust with an emissivity law $\propto$ $\lambda^{-1}$; a single blackbody spectrum reproduces the observed SED very well for each galaxy as seen in figure 3. As a result, we have derived dust temperatures of $23\pm 1$ K for NGC~2841 and $30\pm 2$ K for NGC~2976 (table 4). Although we did not require multiple blackbody components to fit the SEDs, it was clear from figure 2 that NGC~2841 has another component with warmer dust in the nuclear region. If we consider the warmer dust emission occupying about 40 \% of the N60 band flux and remove its contribution from the SED, the dust temperature of NGC~2841 to explain the resultant SED is further reduced down to 21 K, which makes difference in the dust temperature between the galaxies even more conspicuous. 

Figure 4 shows the intensities of the FIS 4 bands between the northwest and southeast ring regions normalized to those in the central region for each galaxy. The intensities in each region are obtained by integrating the surface brightness within an aperture radius of \timeform{0'.5} (as shown by the small dashed circles in figures 1 and 2) without considering aperture corrections. The figure demonstrates that the colors of the far-IR emission from the ring regions are significantly different from that from the center for NGC~2841, while they are quite similar for NGC~2976. Even within the ring structure, the northwest and southwest ring regions have different far-IR colors for NGC~2841, while they are relatively similar for NGC~2976. Hence the dust temperature of NGC~2841 seems to be spatially more variable than that of NGC~2976. 

We have calculated dust masses by using the following equation: 
\begin{equation}
M_{\rm dust} = 5.1\times 10^{-11} S_{\nu} D^2 \lambda^4 (\exp(1.44\times 10^4/(\lambda T_{\rm d}))-1) \hspace{0.5cm}{\rm M_{\odot}},
\end{equation}
where $\lambda$ is a wavelength in $\mu$m, $D$ is a distance in Mpc, and $S_{\nu}$ is a flux density in mJy; the grain emissivity factor is given by Hildebrand (1983) and the average grain radius of 0.1 $\mu$m and the specific dust mass density of 3 g cm$^{-3}$ are assumed. Dust temperatures are set to be equal to those derived from the above SED fitting. Since the HI column densities of both galaxies are no more than several times $10^{22}$ cm$^{-2}$ (Young and Scoville 1982b; Appleton et al. 1981; Simon et al. 2003), the whole of both galaxies are optically thin in far-IR ($\tau_{\rm dust}\lesssim 0.01$ at 60 $\mu$m). The total dust masses thus derived are listed in Table 4. In consequence, we have used these data to derive the same gas-to-dust ratio of $\sim 300$ for both galaxies, which is slightly larger than the accepted value of $100-200$ for our Galaxy (Sodroski et al. 1997).
\subsection{Major axis brightness profile}
The surface brightness profile along the major axis is presented in figure 5 for each of NGC~2841 and NGC~2976, which is obtained by integrating the brightness within $45''$ width across the galactic disk and plotted as a function of the galactocentric distance normalized to the optical major radius for each galaxy (\timeform{4'.1} for NGC2841 and \timeform{3'.0} for NGC2976; de Vaucouleurs et al. 1991). We confirm from the figure that the far-IR distributions have the same extent as the optical disks for both galaxies. Again, we can recognize clear difference in the profile of the N60 bands between the galaxies. The rings of NGC~2841 are located nearer to the galactic center than those of NGC~2976 with respect to the optical size. The positions of the rings do not show any significant change among the FIS bands for each galaxy. 

For NGC~2841, by adopting the distance to the galaxy in table 1, the dust ring is located at radii of about 3.5 kpc, which is well compatible with the molecular/bluer ring at 2-6 kpc (Young and Scoville 1982b; Prieto et al. 1985). However this is a rather cold ($\sim$21 K) dust ring, suggesting that there currently exist no strong star formation activities. As for NGC~2976, the dust ring is located at radii of about 1.4 kpc, which shows excellent spatial coincidence with the H$\alpha$ bright spots (Hodge and Kennicutt 1983). In contrast to NGC~2841, this is a relatively warm ($\sim$30 K) dust ring, which is consistent with a large number of conspicuous H$\alpha$ regions (Hodge and Kennicutt 1983) and thus young concentrated star forming regions contained in the galaxy. By considering a balance between far-IR emission and ultra-violet absorption by thermal dust with the above temperatures and the emissivity factors given by Hildebrand (1983), we have estimated the averaged interstellar radiation field as $G_0 = 5$ for NGC~2841 and $G_0 = 30$ for NGC~2976, where $G_0$ is given in units of the value in the solar neighborhood (1.6$\times 10^{-3}$ ergs cm$^{-2}$ s$^{-1}$; Habing 1968). Hence the difference in the current star formation activity is clear between the galaxies; NGC~2841 is more like our Galaxy. Although molecular rings have hardly been observed in Sc galaxies (Young and Scoville 1982b), we have observed a dust ring structure from the Sc galaxy NGC~2976. The origin of the dust ring might be different between the observed Sb and Sc galaxies; Young and Scoville (1982a) pointed out that the significant difference in the CO distribution our Galaxy and the Sc galaxies is the hole at inner radii rather than the peak at outer radii. The dust ring in the Sb galaxy NGC~2841 could be be due to significant decrease in the amount of the central dust by more rapid star formation in the bulge.
\subsection{Minor axis brightness profile}
Figure 6 shows the surface brightness profile along the minor axis for each of NGC~2841 and NGC~2976, which is obtained by integrating the brightness within $150''$ width and plotted as a function of the galactocentric distance normalized to the optical minor radius for each galaxy (\timeform{1'.8} for NGC2841 and \timeform{1'.4} for NGC2976; de Vaucouleurs et al. 1991). For comparative purposes, the PSF in each FIS band is plotted together in the same horizontal scale as NGC~2976, which is derived by cutting a bright point-like source along the same direction with respect to the scan direction as NGC~2976. It is found from the figure that the far-IR distribution of NGC~2976 is more largely extended than that of NGC~2841 along the direction vertical to the disk as compared with the optical stellar distribution. The extended dust distribution of NGC~2976 might be explained by a relatively shallow gravitational potential for such a low-mass, dwarf-like system as this galaxy (Bronkalla et al. 1992). In addition, there is slight asymmetry in the distribution of NGC~2976, particularly in the N160 and WIDE-L bands, which is attributed to the aforementioned possible protruding emission in the maps of figure 2. This could be the inflow of dust entrained in the intergalactic matter falling to NGC~2976 along the HI bridge; NGC~3077 and M82 in the M81/M82 group suffered close encountours with M81 during the last $10^9$ years (Cottrell 1977; Brouillet et al. 1991), generating the huge cloud linking NGC~2976 to the central group by tidal interactions (Appleton et al. 1981; Appleton and van der Hulst 1988). Therefore the infall gas may have already been enriched with dust, supporting the possible detection of the dust inflow, which is also compatible with the above result that the gas-to-dust ratio of NGC~2976 is not so large but similar to that of NGC~2841. 
Hence the recent star forming activity and the formation of the dust ring in NGC~2976 may have been influenced from outside, triggered by interaction with the intergalactic cloud. Another possible explanation for the minor-axis dust asymmetry in NGC~2976 is that the galaxy has experienced outflows similar to that of M82. In contrast, NGC~2841 shows relatively symmetrical far-IR distributions across the minor axis, which is consistent with a prediction by Elmegreen and Block (1999) who have demonstrated that dust embedded in an inclined exponential disk can produce such strong minor-axis color gradient as observed from NGC~2841 (Block et al. 1996), even if the dust is distributed symmetrically.  
\subsection{Dimension analysis of variation in SED}
In order to estimate the number of independent components involved in the surface brightness variation within the galaxy, we have performed a dimension analysis. The FIS 4-band images are first rebinned with the common spatial scale of $15''\times 15''$. The brightness in each spatial bin is evaluated in the four bands, so we can define a four-dimensional vector $\vec{v_i}$ for the $i$-th spatial pixel. The index $i$ runs over the field where a signal-to-noise ratio exceeds 5 in the WIDE-L band, and each component of the vector is normalized to the peak in each band. If the brightness variation involves no SED changes, the four components will change proportionally to one another, so that the vectors \{$\vec{v_i}$\} all trace the same line in the four-dimensional vector space. If instead the distribution of dust emission is composed of two spectrally distinct components which fluctuate independently, the set \{$\vec{v_i}$\} will define a two-dimensional figure in the vector space. In practice, the calculation can be performed in the following way: let the brightness in the $i$-th bin and the $k$-th photometric band be $x_i^k$. Then, the distance, $d_{ij}$ of the two vectors, $\vec{v_i}$ and $\vec{v_j}$, is calculated to be
\begin{equation}
d_{ij}=\sqrt{\sum_{k=1}^{4}(x_i^k-x_j^k)^2}.
\end{equation}  
The number of combinations of the two vectors satisfying the condition of $d_{ij}<r$ is defined as $N(r)$. If the set \{$\vec{v_i}$\} defines a $p$-dimensional figure, $N(r)$ is expected to increase in proportion to $r^p$ ($r\leq 4$). 

Figure 7 shows the curves $N(r)$ calculated for NGC~2841 and NGC~2976; $N(r)$ saturates at large values of $r$ due to the limited number of the vectors, while $N(r)$ is dominated by noise at small $r$'s, and expected to yield $p\sim 4$ because the noise in different bands are independent. As seen in the figure, both curves approximately following the power-law index of 1, the slope for NGC~2841 is significantly steeper than that for NGC~2976 at intermediate $r$'s. Hence it is numerically demonstrated that the temperature of dust in NGC~2841 is spatially more variable than that in NGC~2976 within the galaxy. 
\section{Conclusions}
We have presented new far-IR images of nearby nearly-edge-on Sb galaxy NGC2841 and Sc galaxy NGC2976 at 65, 90, 140, and 160 $\mu$m taken with AKARI/FIS. Both galaxies have revealed similar dust ring morphologies. The property of the dust ring, however, is quite different between the galaxies: a cold dust ring with rather variable temperatures for NGC~2841 and a warm dust ring with relatively uniform temperatures for NGC~2976, which presumably reflects the difference in the origin of the ring structure for the different Hubble type of the galaxy. In addition to the dust ring, a warm dust component is detected from the center of NGC~2841, which may be attributed to the heating by its LINER nucleus. As for NGC~2976, an extended dust component is observed along the minor axis, which includes the possible emission protruding to one side from the disk; this might indicate dust inflow along the HI bridge that connects NGC~2976 to the huge cloud enshrouding the members of the M81/82 group. To further this work, we await future far-IR observations with higher spatial resolution and sensitivity by space missions such as Herschel (Pilbratt 2004) and SPICA (Kaneda et al. 2004; Onaka et al. 2005; Nakagawa and Murakami 2007). 

\bigskip

We thank all the members of the AKARI project, particularly those deeply engaged in the observation planning and the satellite operation during the performance verification phase. We would also express our gratitude to the AKARI data reduction team for their dedicated work in generating the TSD and developing data analysis pipelines.  {\it AKARI} is a JAXA project with the participation of ESA. This work is financially supported by a Grant-in-Aid from the Ministry of Education, Culture, Sports, Science and Technology in Japan (No. 17740123).

\clearpage

\begin{figure}
\FigureFile(65mm,65mm){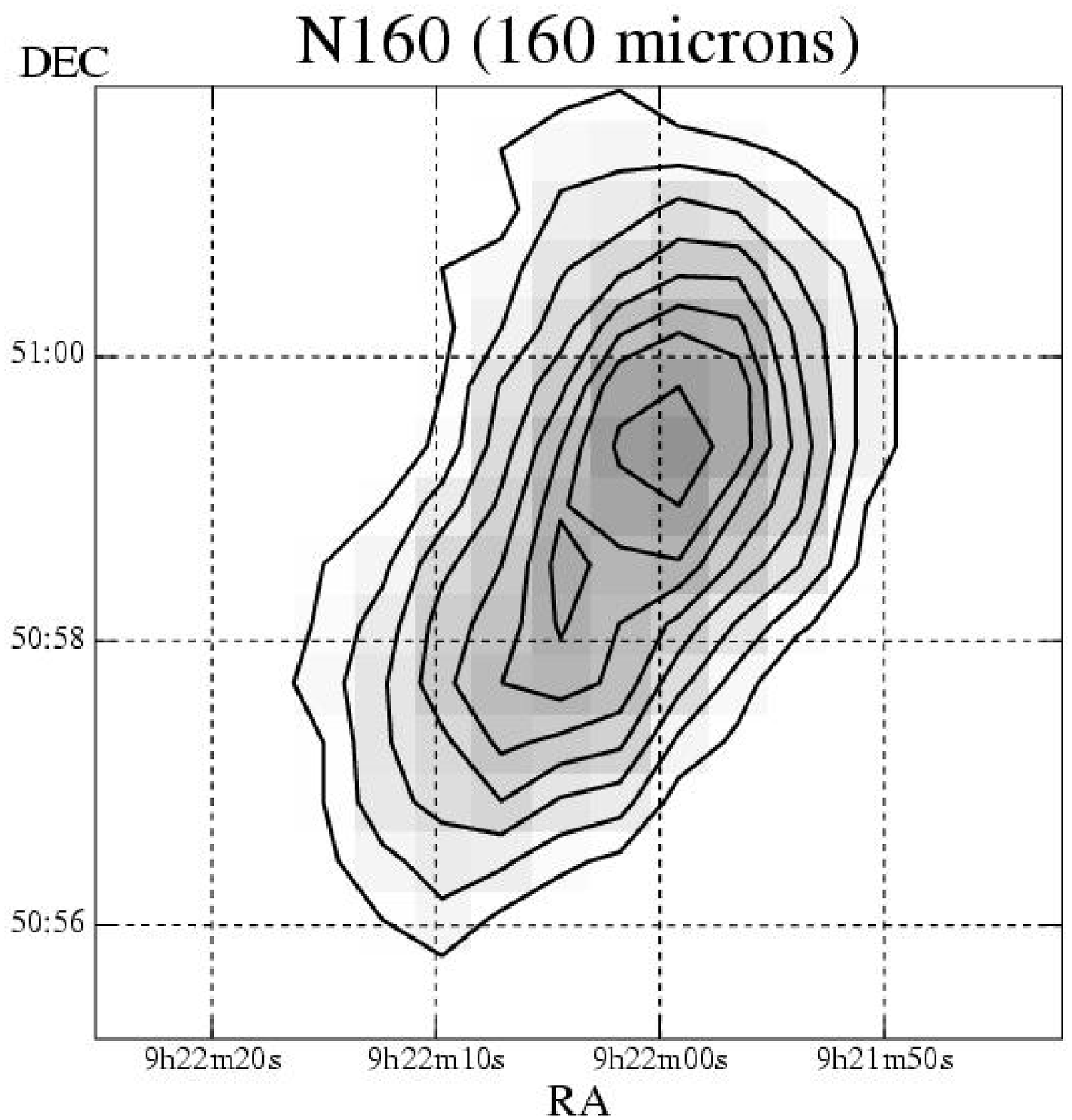}\FigureFile(65mm,65mm){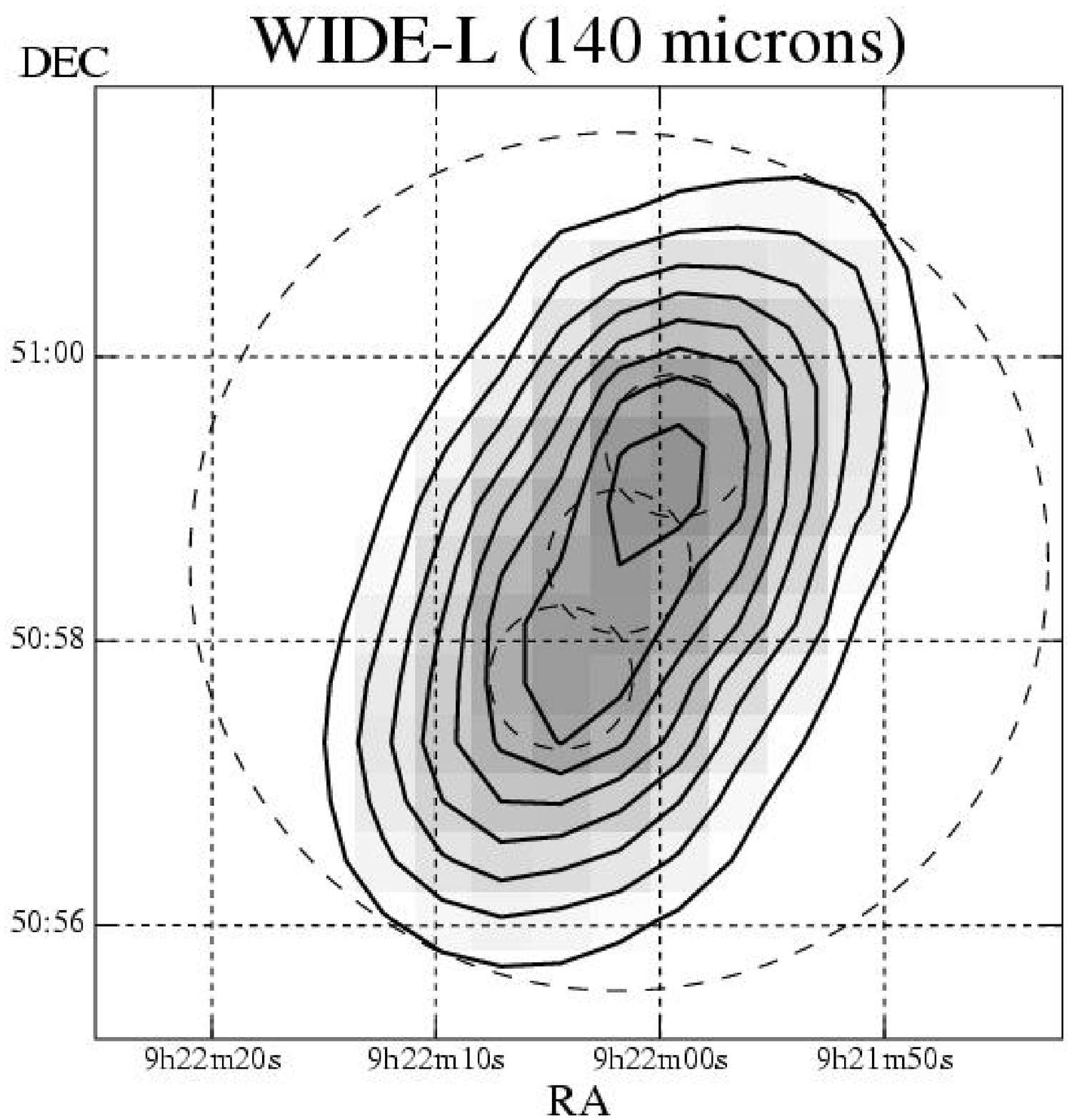}\\
\FigureFile(65mm,65mm){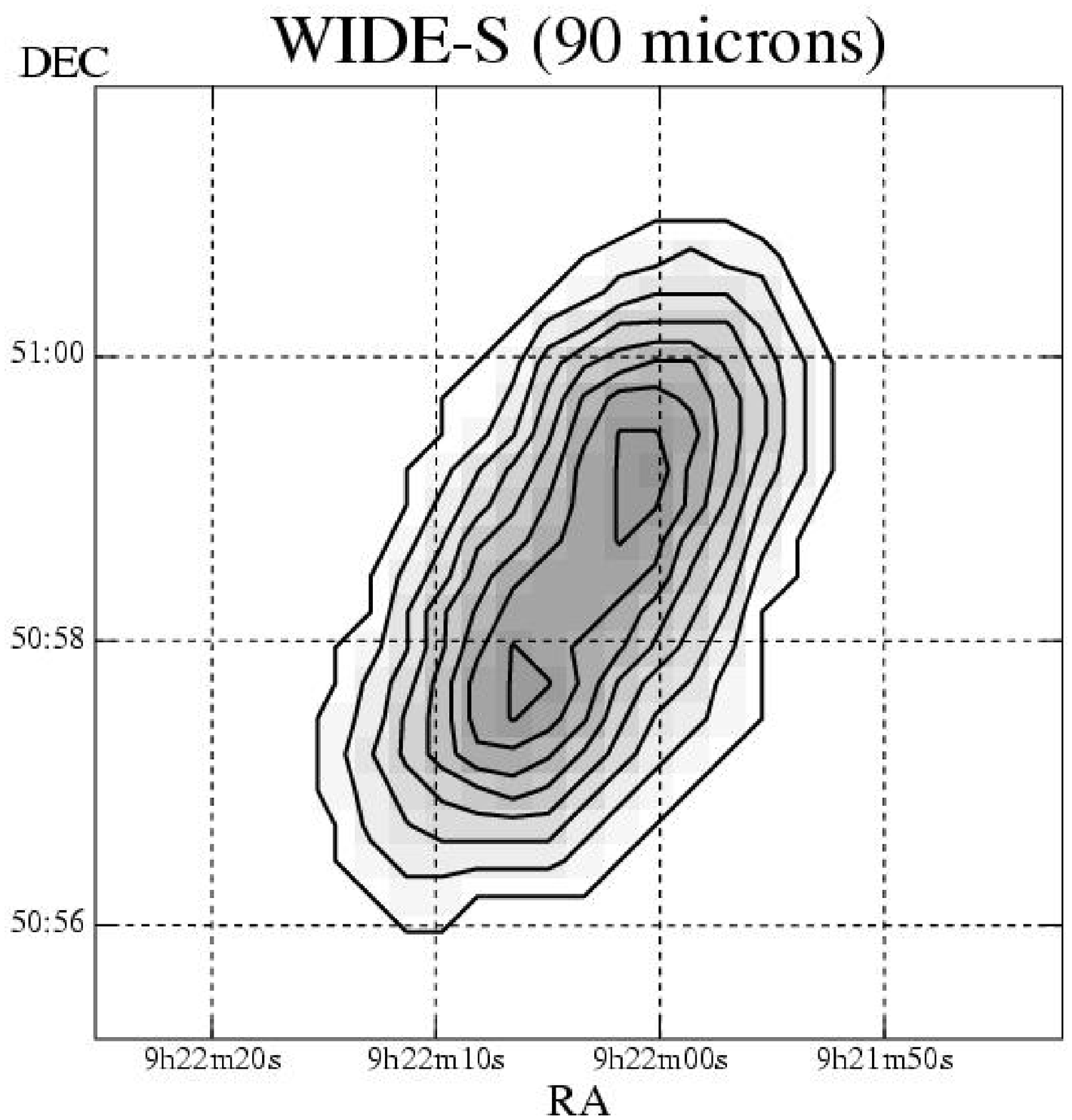}\FigureFile(65mm,65mm){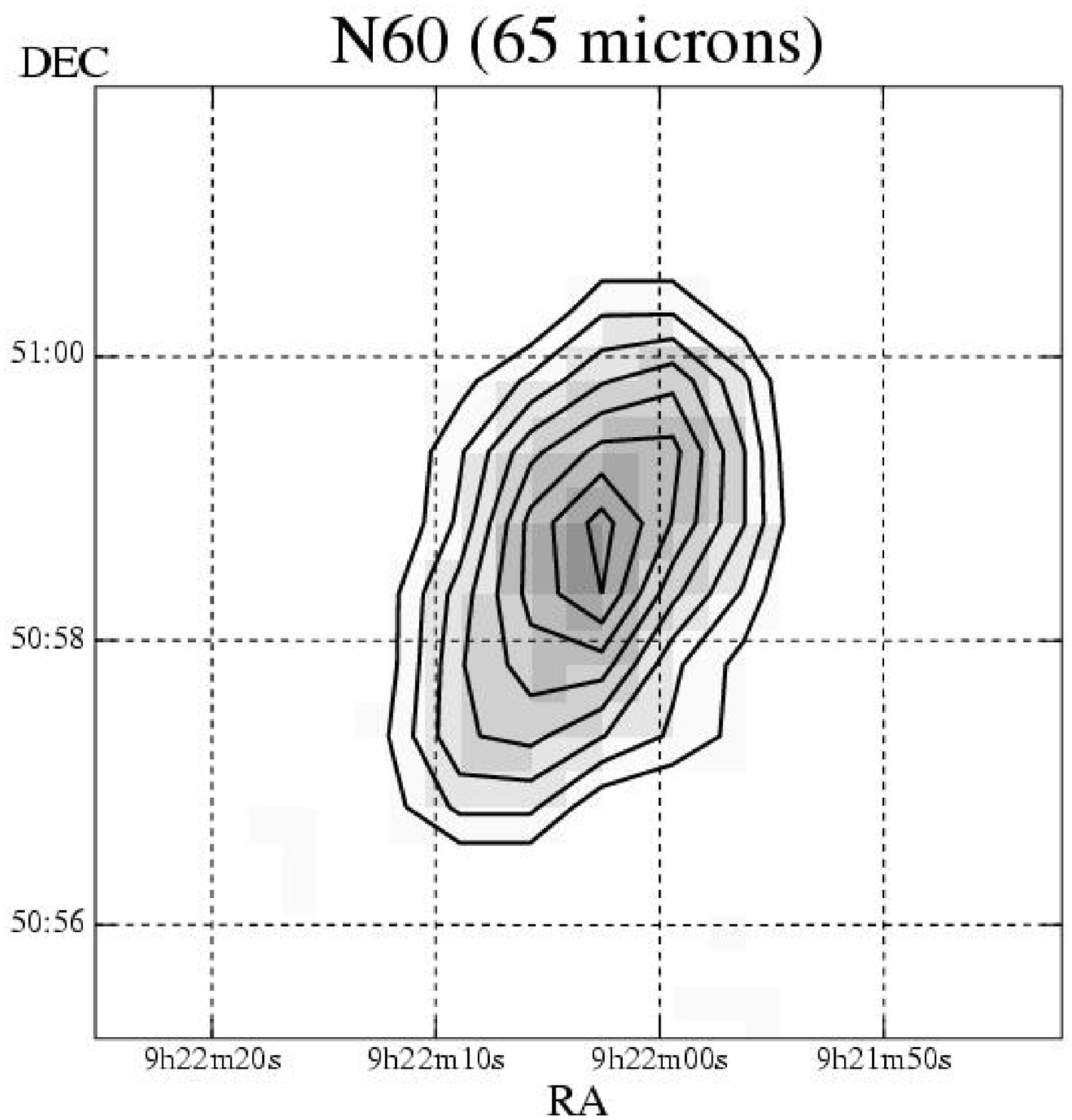}\\
\caption{FIS 4-band images of NGC~2841. For each panel, contours are drawn from 25 \% to 95 \% levels with the linear step of 10 \% of the background-subtracted peak surface brightness of 58 MJy/str (N160), 67 MJy/str (WIDE-L), 24 MJy/str (WIDE-S), and 28 MJy/str (N60). The dashed circles in the WIDE-L image correspond to photometry apertures, for which the flux densities in table 3 and figure 3 as well as the relative band intensities in figure 4 are derived.}
\end{figure}
\clearpage

\begin{figure}
\FigureFile(65mm,65mm){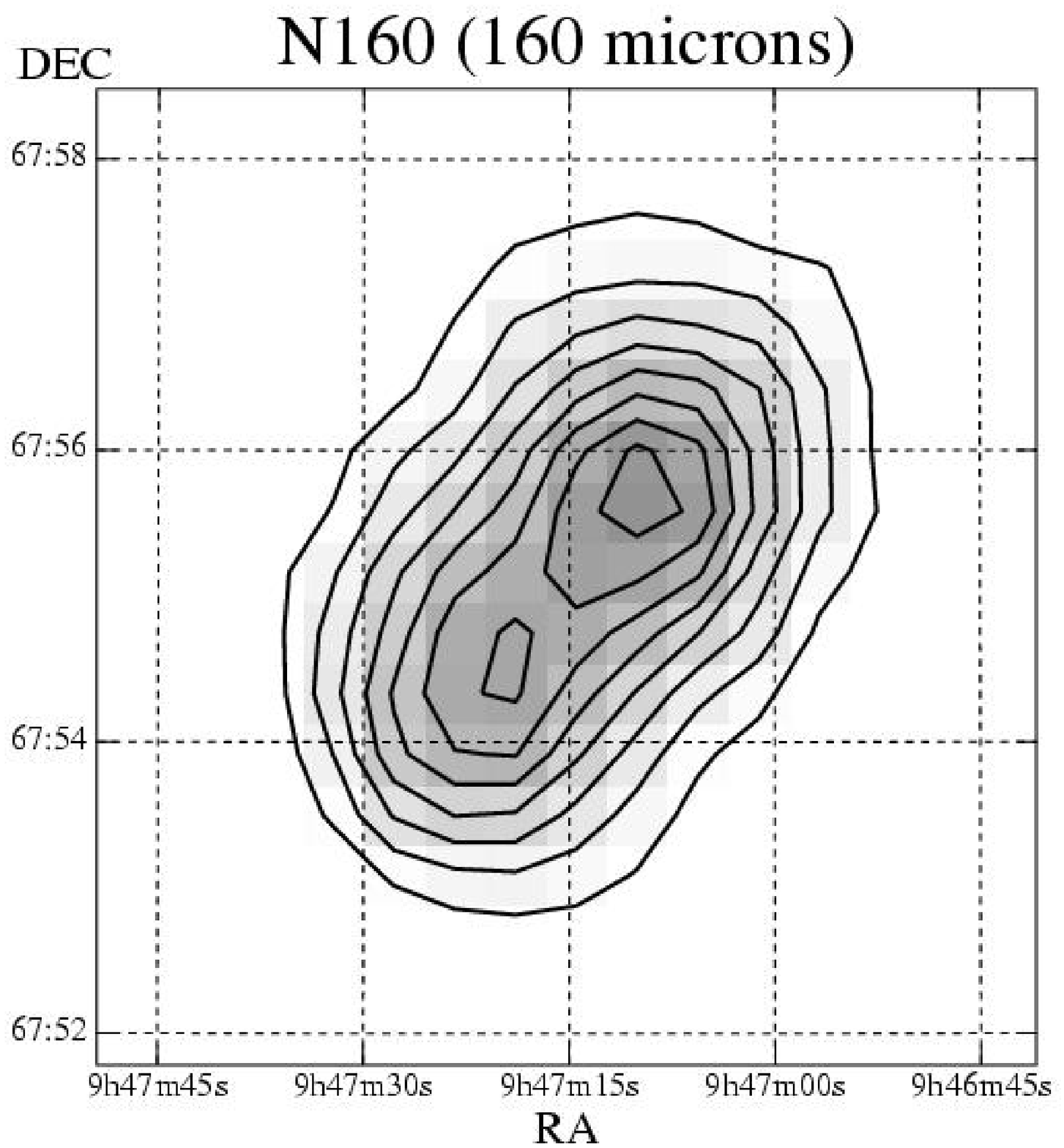}\FigureFile(65mm,65mm){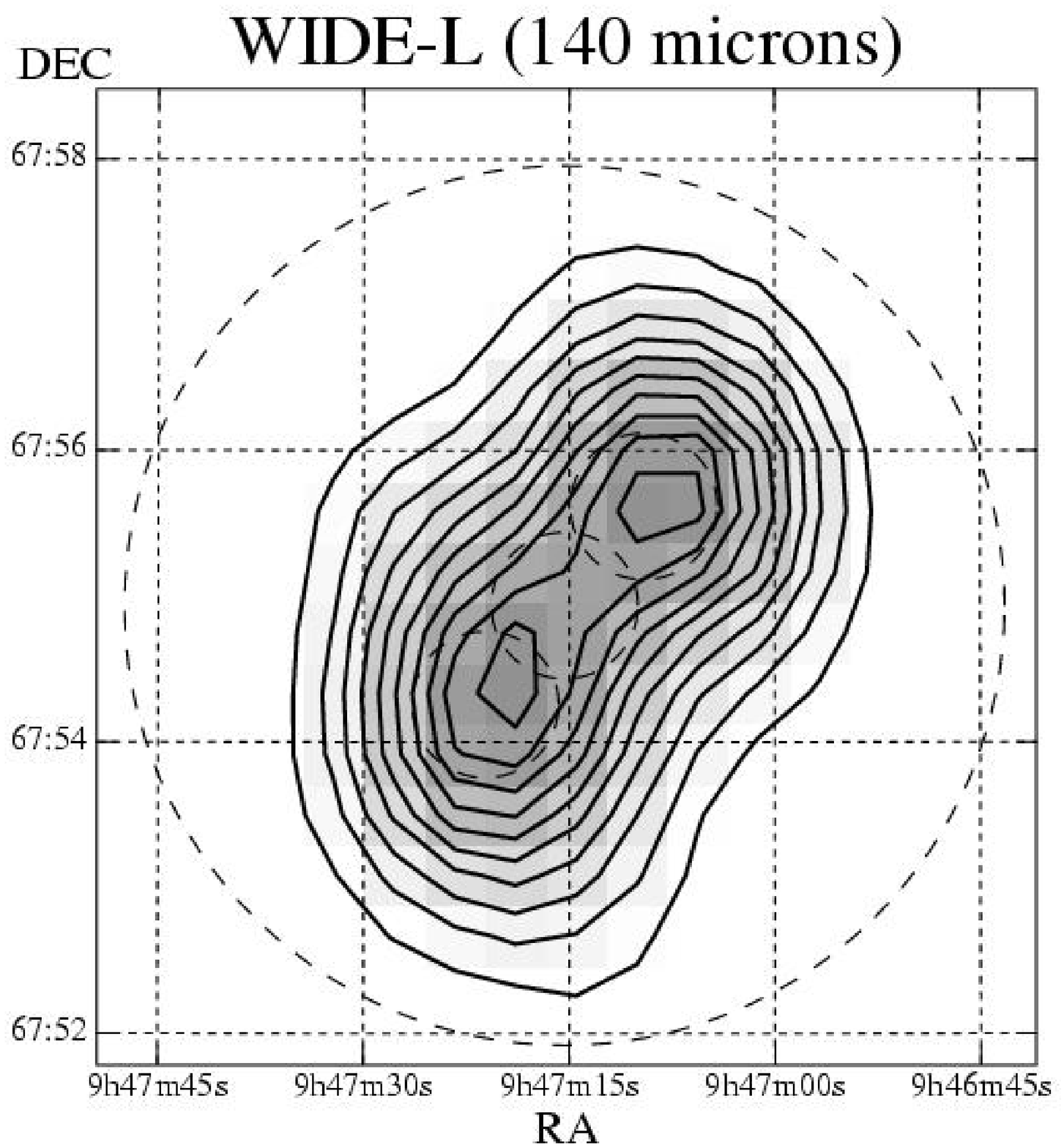}\\
\FigureFile(65mm,65mm){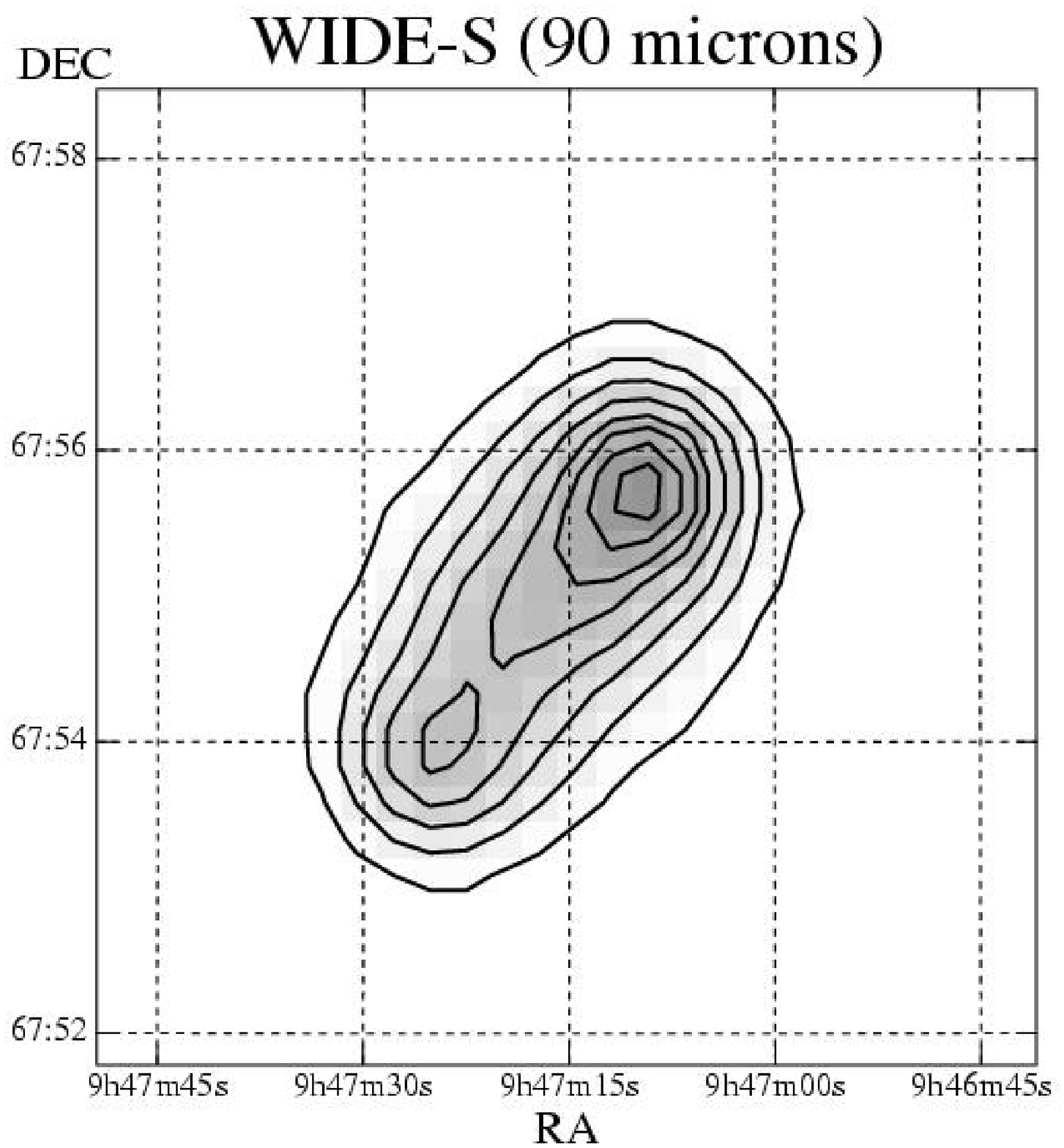}\FigureFile(65mm,65mm){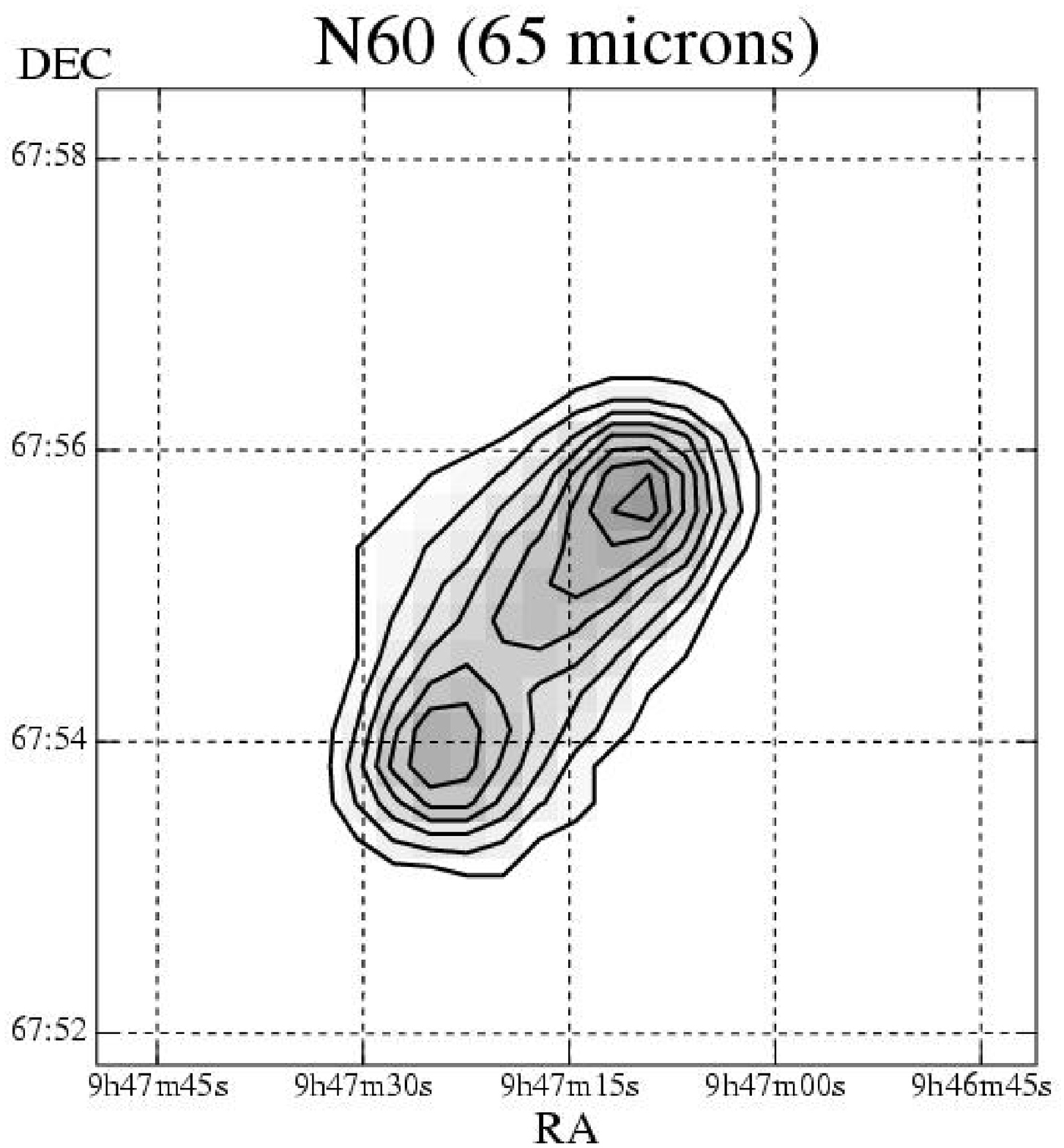}\\
\caption{Same as Fig.1, but for NGC~2976. The background-subtracted peak surface brightness is 82 MJy/str (N160), 79 MJy/str(WIDE-L), 77 MJy/str (WIDE-S), and 62 MJy/str (N60).}
\end{figure}
\clearpage

\begin{figure}
\FigureFile(85mm,85mm){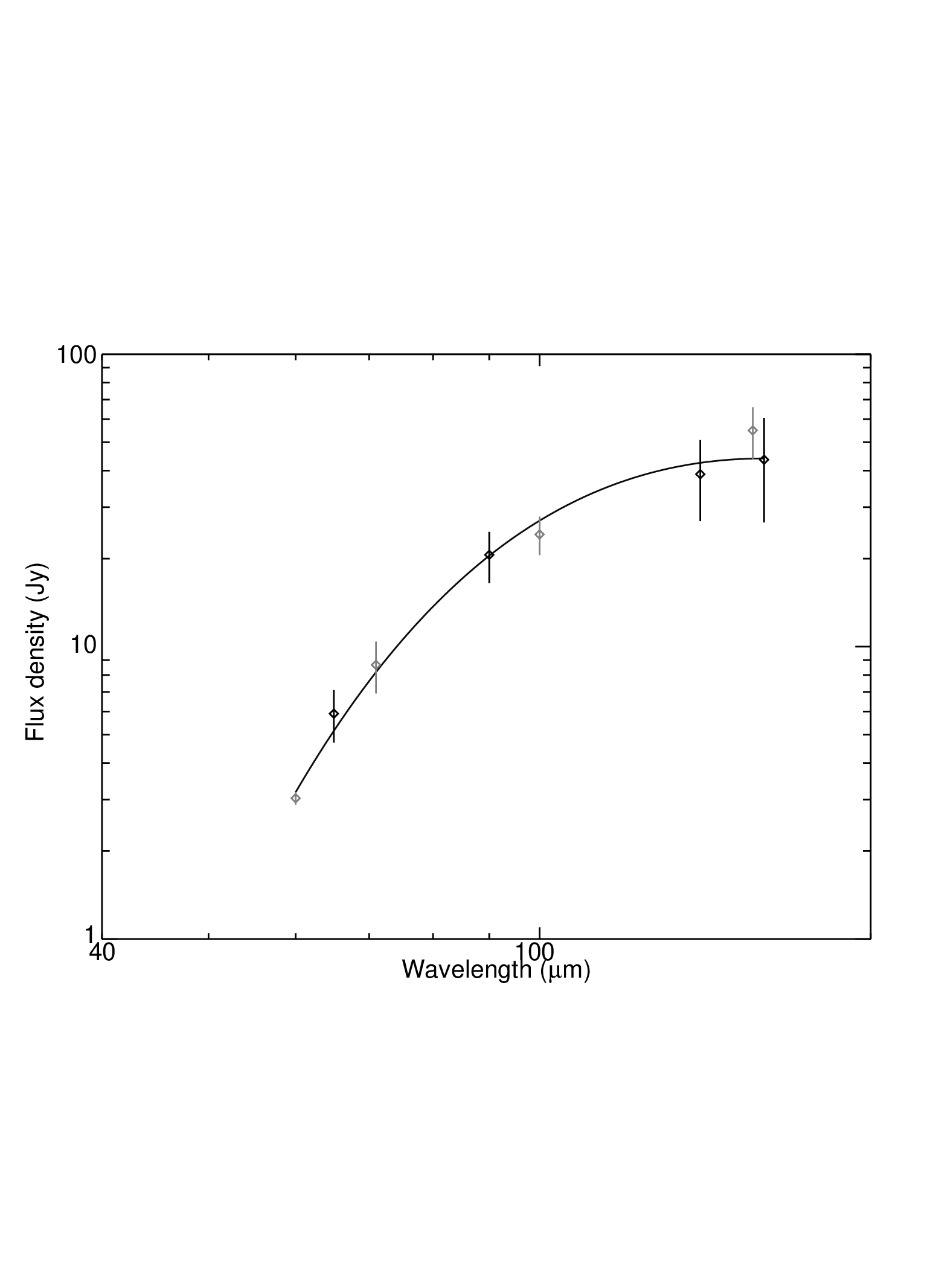}\FigureFile(85mm,85mm){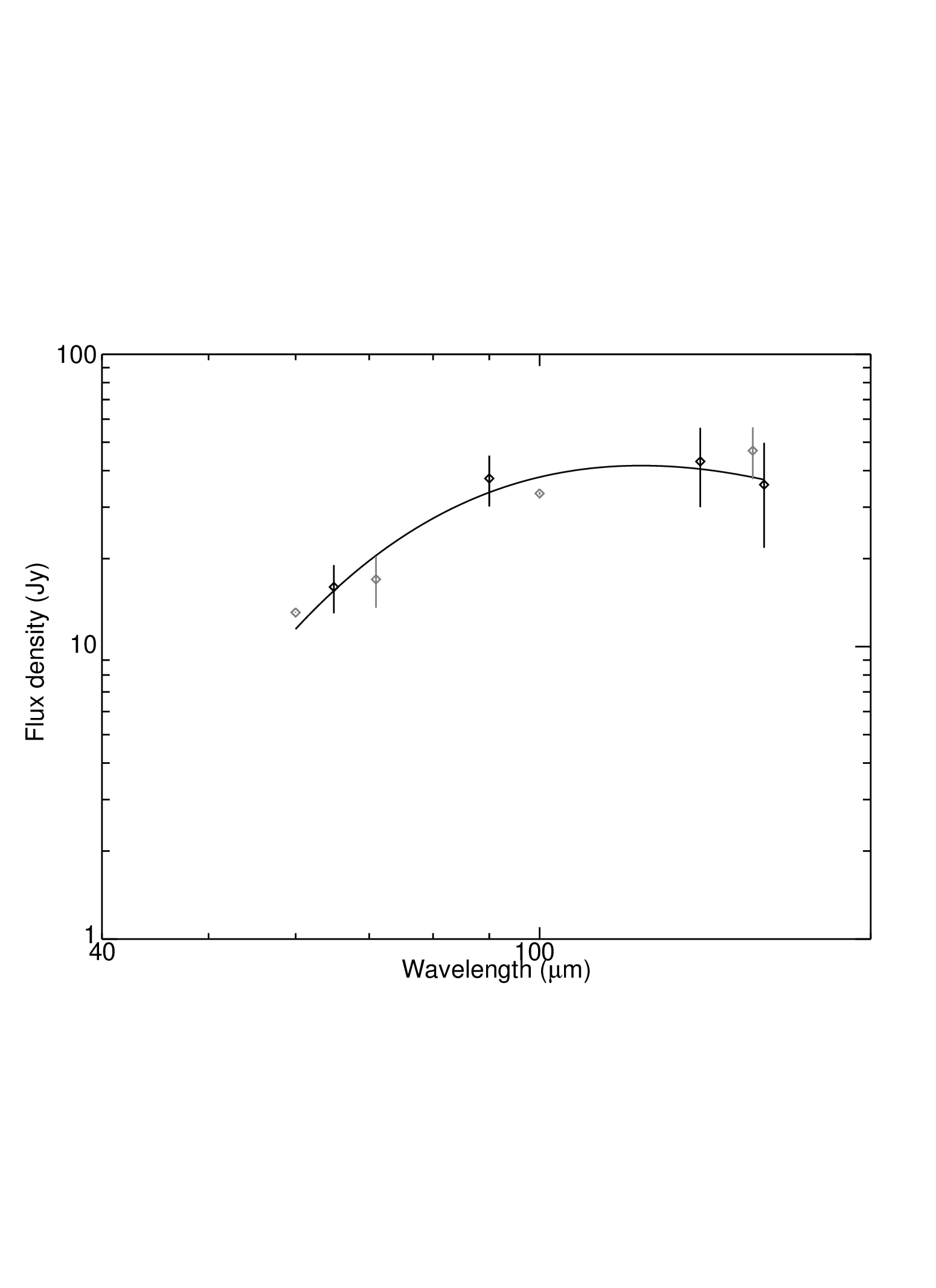}
\caption{Spectral energy distributions of NGC~2841 (left) and NGC~2976 (right) obtained with the FIS 4 bands (in black) together with the data in the IRAS 60 $\mu$m, IRAS 100 $\mu$m, Spitzer/MIPS 70 $\mu$m, and MIPS 160 $\mu$m bands (in gray; table 3).  The flux densities of the FIS are obtained by integrating the brightness within an aperture radius of 3' as shown in figures 1 and 2. Blackbody spectra (emissivity law $\propto$ $\lambda^{-1}$) fitted to the SEDs with temperatures of 23 K and 30 K are plotted together for NGC~2841 and NGC~2976, respectively.}
\end{figure}
\clearpage

\begin{figure}
\FigureFile(85mm,85mm){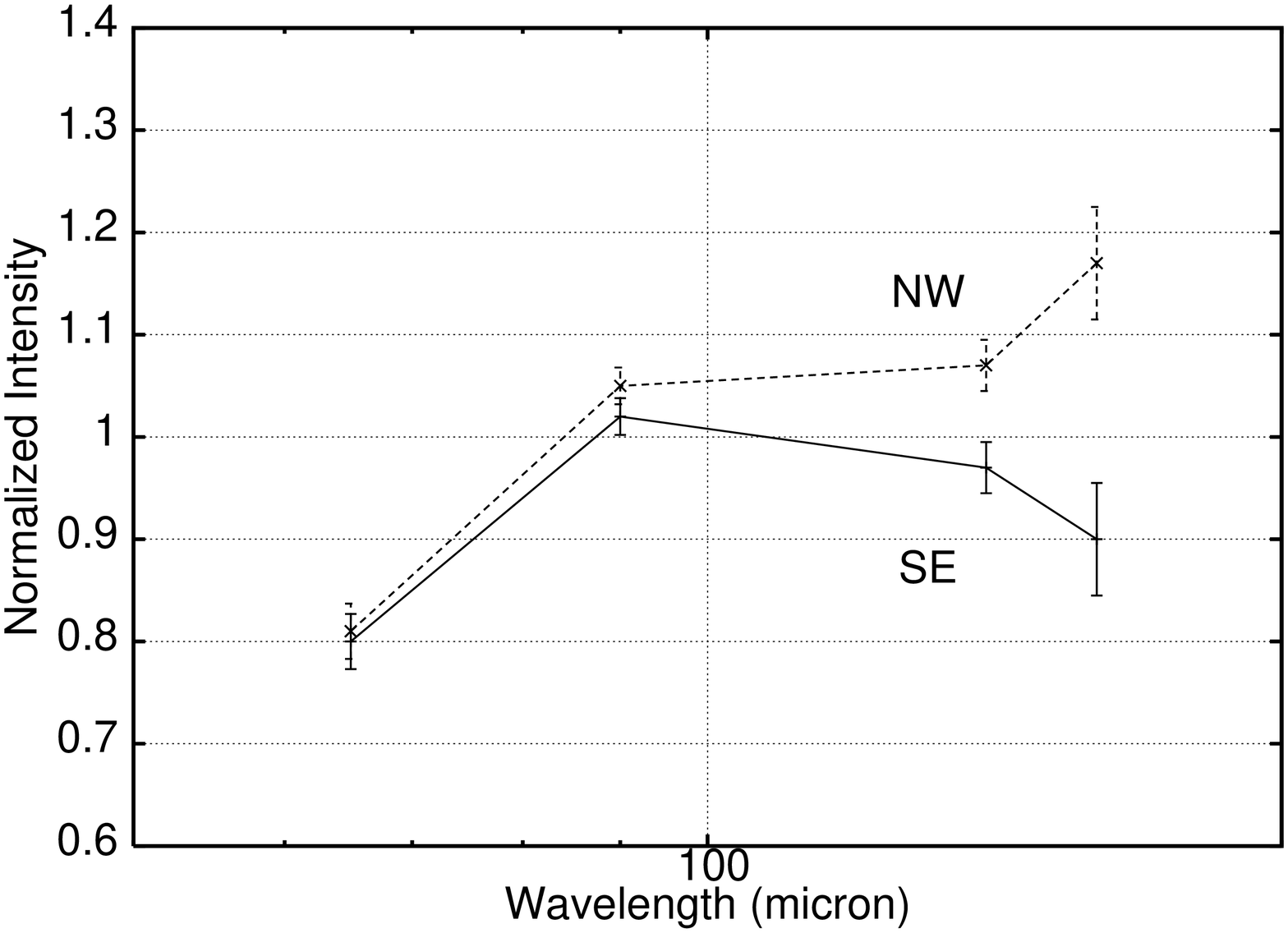}\FigureFile(85mm,85mm){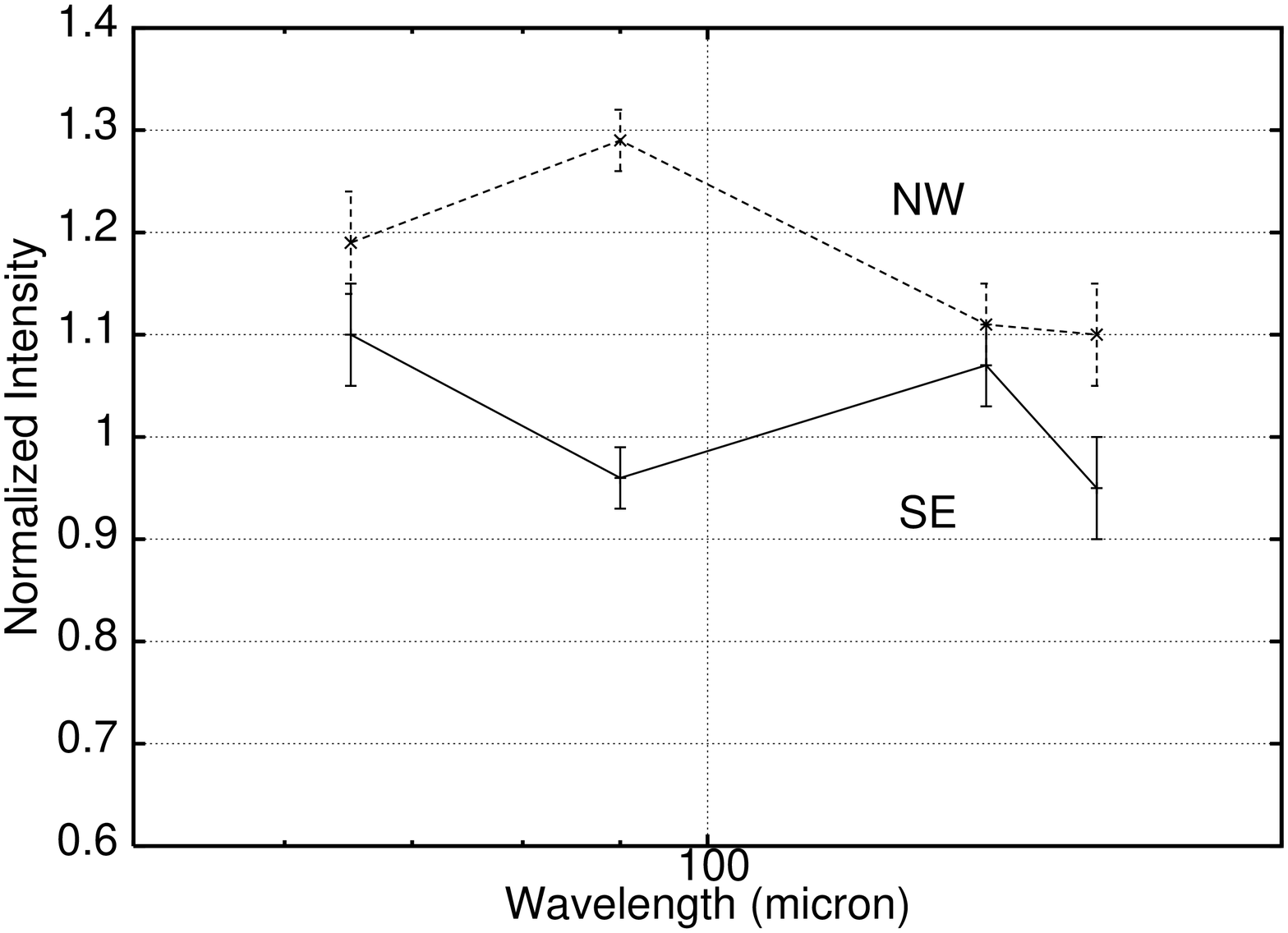}
\caption{Intensities of the FIS 4 bands in the northwest and southeast ring regions of NGC~2841 (left) and NGC~2976 (right) normalized to those in the central regions of the galaxies. The intensities in each region are obtained by integrating the brightness within an aperture radius of \timeform{0'.5} as shown in figure 1. }
\end{figure}
\clearpage

\begin{figure}
\FigureFile(120mm,120mm){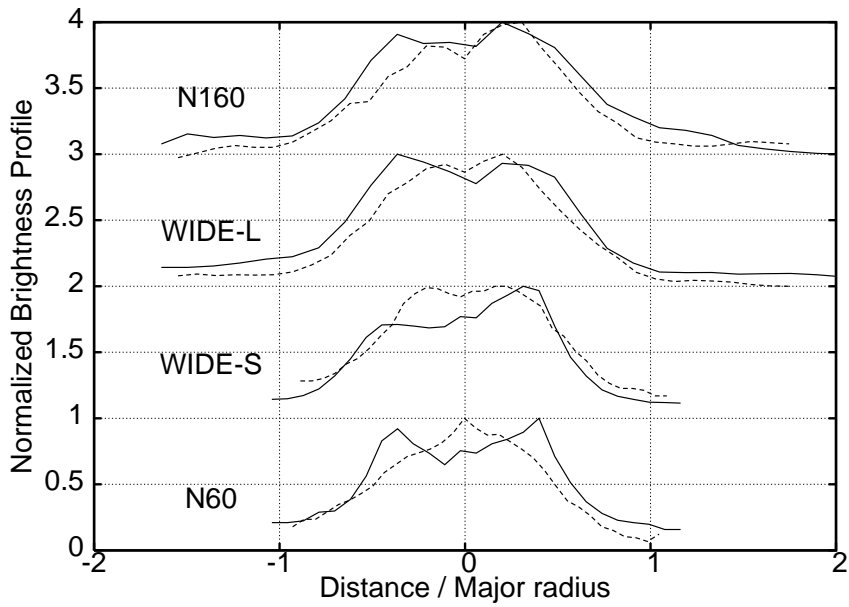}
\caption{Major-axis brightness profiles of NGC~2841 (dashed lines) and NGC~2976 (solid lines) in the FIS 4 bands. These are obtained by integrating the brightness within $45''$ width of the galactic disks and plotted as a function of the galactocentric distances normalized to the optical major radii, \timeform{4'.1} for NGC~2841 and \timeform{3'.0} for NGC~2976. Each profile is normalized to its peak and shifted along the vertical axis for clarity of presentation.} 

\end{figure}
\clearpage

\begin{figure}
\FigureFile(120mm,120mm){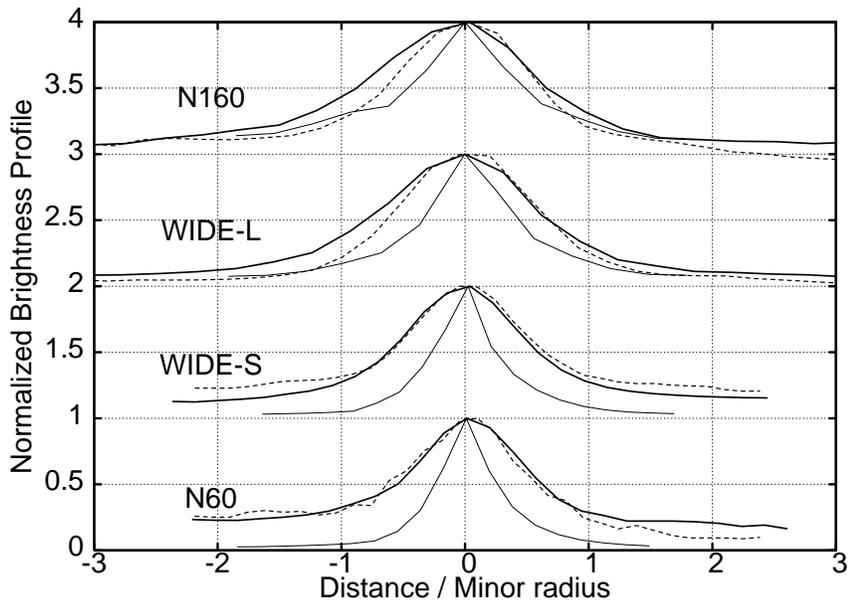}
\caption{Minor-axis brightness profiles of NGC~2841 (dashed lines) and NGC~2976 (thick solid lines) in the FIS 4 bands. These are obtained by integrating the brightness within $150''$ width along the galactic disks and plotted as a function of the galactocentric distances normalized to the optical minor radii, \timeform{1'.8} for NGC~2841 and \timeform{1'.4} for NGC~2976. The point spread function of each band is plotted together (thin solid line), as compared with the profile of NGC~2976. Each profile is normalized to its peak and shifted along the vertical axis for clarity of presentation.} 

\end{figure}
\clearpage

\begin{figure}
\FigureFile(120mm,120mm){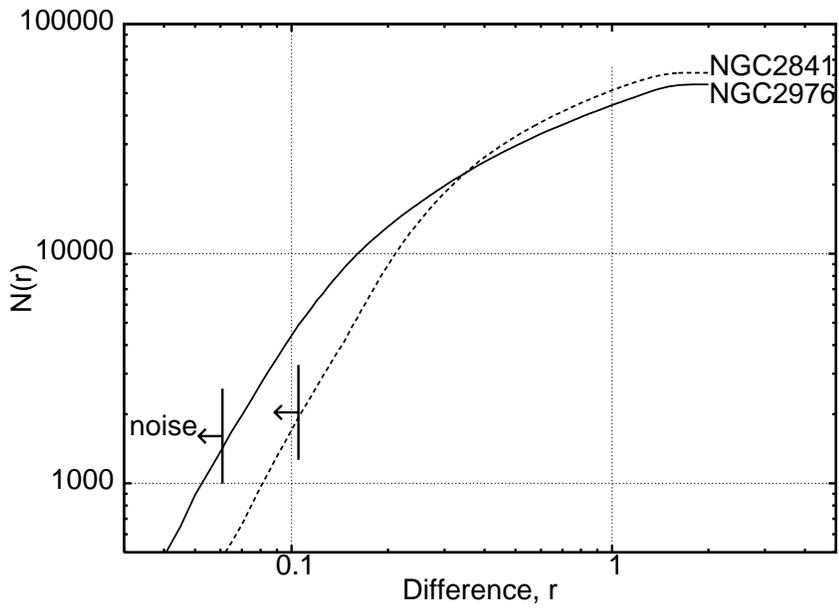}
\caption{Results of the dimension analysis (see text for details). The horizontal axis, $r$, corresponds to the difference of the two 4-dimensional vectors, which are constructed from the FIS 4-band intensities normalized to the peak of each band, while the vertical axis, $N(r)$, is the number of combinations of the two vectors whose difference is less than $r$. } 
\end{figure}

\clearpage

\begin{table*}
\caption{Properties of the Galaxies}
\begin{center}
\begin{tabular}{lccc}
%\rotate
\hline\hline
  &&NGC~2841&NGC~2976\\ \hline
Type&:&Sb&Sc\\
Distance\footnotemark[$*$]&:&7.6 Mpc&3.6 Mpc\\
Inclination\footnotemark[$\dagger$]&:&$68^{\circ}$&$61^{\circ}$\\
Position angle\footnotemark[$\dagger$]&:&$148^{\circ}$&$142^{\circ}$\\
Optical size\footnotemark[$\ddagger$]&:&\timeform{8'.1}$\times$\timeform{3'.5}&\timeform{5'.9}$\times$\timeform{2'.7}\\
Luminosity\footnotemark[$\S$]&:&$2\times 10^{10}$L$_{\odot}$&$4\times 10^{8}$L$_{\odot}$\\
Total mass\footnotemark[$\S$]&:&$7\times 10^{10}$M$_{\odot}$&$3\times 10^{9}$M$_{\odot}$\\
HI mass\footnotemark[$\|$]&:&$5\times 10^{8}$M$_{\odot}$&$6\times 10^{7}$M$_{\odot}$\\
H$_{2}$ mass\footnotemark[$\|$]&:&$9\times 10^{8}$M$_{\odot}$&$4\times 10^{7}$M$_{\odot}$\\
\hline
\\
\multicolumn{4}{@{}l@{}}{\hbox to 0pt{\parbox{160mm}{\footnotesize

\par\noindent
\footnotemark[$*$] Osman et al. (1982) for NGC~2841 and Karachentsev et al. (2002) for NGC~2976  
\par\noindent
\footnotemark[$\dagger$] Bosma (1978) for NGC~2841, and Appleton et al. (1981) and Bronkalla et al. (1992) for NGC~2976 
\par\noindent
\footnotemark[$\ddagger$] de Vaucouleurs et al. (1991) 
\par\noindent
\footnotemark[$\S$] Young \& Scoville (1982b) for NGC~2841 and Carozzi-Meyssonnier (1980) for NGC~2976
\par\noindent
\footnotemark[$\|$] Young \& Scoville (1982b) for NGC~2841 and Brouillet et al.(1992) for NGC~2976
}\hss}}
\end{tabular}
\end{center}
\end{table*}

\clearpage

\begin{table*}
\caption{Observation Log}
\begin{center}
\begin{tabular}{lccc}
%\rotate
\hline\hline
  &&NGC~2841&NGC~2976\\ \hline
R.A. (J2000)&:&\timeform{9h22m02s.63}&\timeform{9h47m15s.46}\\
Decl. (J2000)&:&\timeform{50D58'35''.5}&\timeform{67D54'59''.0}\\
Date&:&Apr 26, 2006&Apr 20, 2006\\
Observing mode&:&FIS01&FIS01\\
Scan speed&:&$8''$/sec&$8''$/sec\\
Reset interval&:&0.5 sec&0.5 sec\\ \hline
\end{tabular}
\end{center}
\end{table*}

\clearpage

\begin{table*}
\caption{Flux Densities of the Galaxies}
\begin{center}
\begin{tabular}{lccc}
%\rotate
\hline\hline
  &&NGC~2841&NGC~2976\\ \hline
FIS N160&:&44$\pm$17 Jy&36$\pm$14 Jy\\
FIS WIDE-L&:&39$\pm$12 Jy&43$\pm$13 Jy\\
FIS WIDE-S&:&20.6$\pm$4.1 Jy&37.6$\pm$7.4 Jy\\
FIS N60&:&$5.9\pm$1.2 Jy&16.0$\pm$3.2 Jy\\
MIPS 160\footnotemark[$*$]&:&55$\pm$11 Jy&46.8$\pm$9.4 Jy\\
MIPS 70\footnotemark[$*$]&:&8.7$\pm$1.7 Jy&17.0$\pm$3.4 Jy\\
IRAS 100\footnotemark[$\dagger$]&:&24.2$\pm$3.6 Jy&33.43$\pm$0.35 Jy\\
IRAS 60\footnotemark[$\dagger$]&:&3.03$\pm$0.15 Jy&13.09$\pm$0.03 Jy\\ 
\hline
\\
\multicolumn{4}{@{}l@{}}{\hbox to 0pt{\parbox{160mm}{\footnotesize

\par\noindent
\footnotemark[$*$] Dale et al. (2005) for both NGC~2841 and NGC~2976  
\par\noindent
\footnotemark[$\dagger$] Rice et al. (1988) for NGC~2841 and Sanders et al. (2003) for NGC~2976 
}\hss}}

\end{tabular}
\end{center}
\end{table*}

\clearpage

\begin{table*}
\caption{Derived Properties of Dust in the Galaxies}
\begin{center}
\begin{tabular}{lccc}
%\rotate
\hline\hline
  &&NGC~2841&NGC~2976\\ \hline
$T_{\rm dust}$&:&23$\pm$1 K&30$\pm$2 K\\
$M_{\rm dust}$&:&$4\times 10^6$ M$_{\odot}$&$3\times 10^5$ M$_{\odot}$\\
$M_{\rm gas}$/$M_{\rm dust}$\footnotemark[$*$]&:&300&300\\
\hline
\\
\multicolumn{4}{@{}l@{}}{\hbox to 0pt{\parbox{160mm}{\footnotesize

\par\noindent
\footnotemark[$*$] Total gas mass is derived from table 1.
}\hss}}

\end{tabular}
\end{center}
\end{table*}

\end{document}